# Towards magnetic slowing of atoms and molecules

E. Narevicius, C. G. Parthey, A. Libson, M. F. Riedel, U. Even, and M. G. Raizen

We outline a method to slow paramagnetic atoms or molecules using pulsed magnetic fields. We also discuss the possibility of producing trapped particles by adiabatic deceleration of a magnetic trap. We present numerical simulation results for the slowing and trapping of molecular oxygen.

The control of atomic and molecular motion has been a long-standing scientific goal with many potential technological applications ranging from atomic lithography to matter-wave interferometry for inertial sensing[1]. The standard method to date for controlling atomic motion has been laser cooling. Although this approach is very successful, it has been limited to the alkalis, alkaline earths, meta-stable noble-gas atoms and a few others species. Most species of technological interest (such as ferromagnetic atoms and semiconductor atoms) are not amenable to laser cooling due to their complex internal structure.

Other methods that do not rely on laser cooling include the pulsed electric field slower for polar molecules[2], the pulsed optical field slower[3], and crossed molecular beam slowing by billiard like collisions[4]. Mechanical control methods include a spinning supersonic source[5], a moving magnetic mirror[6] and the slowing of helium atoms by elastic reflection from a receding crystal mounted on a rotor[7,8].

We consider in this paper a new and quite general approach for creating slow cold atom and molecular beams: atoms will be slowed in a time-dependent magnetic field where low-field seekers are forced to climb a magnetic hill, losing kinetic energy in the process.

In this paper, we outline the basic elements of the proposed method and indicate our experimental progress towards this goal with realistic parameters. We start from a description of the supersonic beam. We discuss the direct magnetic deceleration of atoms, followed by the concept of a moving magnetic trap that is decelerated to adiabatically slow trapped atoms or molecules. We finally discuss possible applications of this method.

## Supersonic Beams

Atomic beams are traditionally created by allowing gas to escape from a source with a small aperture and collimating the output. In the dilute-gas regime, where the mean free path is larger than the aperture size, the resulting beam has a very broad distribution in velocity as well as angle. As the pressure is increased in the source the resulting beam becomes very monochromatic and directional. Supersonic expansion of the atoms as they escape from the aperture leads to these properties, and such beams have become important tools in physical chemistry[9]. To reach the required pressures of several atmospheres, noble gases are typically used as the primary gas and they are "seeded" with another gas which is carried along. It is possible to seed almost any element into the supersonic beam using the process of laser ablation.

The characteristics of the beam are critical for the proposed slowing and the parameters in this paper are based on a pulsed supersonic nozzle developed by Even et al.[10,11] that offers a pulse duration as short as 10 μs FWHM with a repetition rate of 40 Hz, combined with cryogenic operation. The resulting beam has a brightness of 4 x $10^{23}$ atoms/sr/sec, an order of magnitude better than any source reported to date.

**Magnetic Deceleration of Atoms**

We propose that paramagnetic atoms can be slowed using a pulsed magnetic field. This is in a complete analogy with the pulsed electric field decelerator which has been used to slow polar molecules. Most atoms in the periodic table have a permanent magnetic moment allowing the control of the atomic motion using magnetic fields. An early example of such control is the splitting of an atomic silver beam in a Stern-Gerlach experiment[12]. The principle of magnetic deceleration is conceptually simple: low field seekers lose kinetic energy by moving into the high magnetic field region at the center of the electromagnetic coil. When the atom reaches the top of the magnetic "hill" the magnetic field is suddenly switched off. Due to conservation of energy the amount of the kinetic energy lost is equal to the Zeeman energy shift[13],

$$\Delta \mathrm{E} = g_s \mu_\beta M_J H$$

where $g_s$ is the Landé factor, $\mu_\beta$ is the Bohr magneton, $M_J$ is the projection of the total angular momentum on the quantization axis and H is the magnetic field.

The key question is whether short pulses of large magnetic field can be generated to stop an atomic beam over a distance of less than 1 meter. We have developed a miniature electromagnetic coil, which is able to deliver 2-3 Tesla magnetic fields throughout a 100 µsec pulse, according to our calculations. Based on our simulations, we predict that a supersonic molecular beam of $O_2$ can be slowed from 250 m/s to 50 m/s using 54 pulsed magnetic stages that span 76 centimeters.

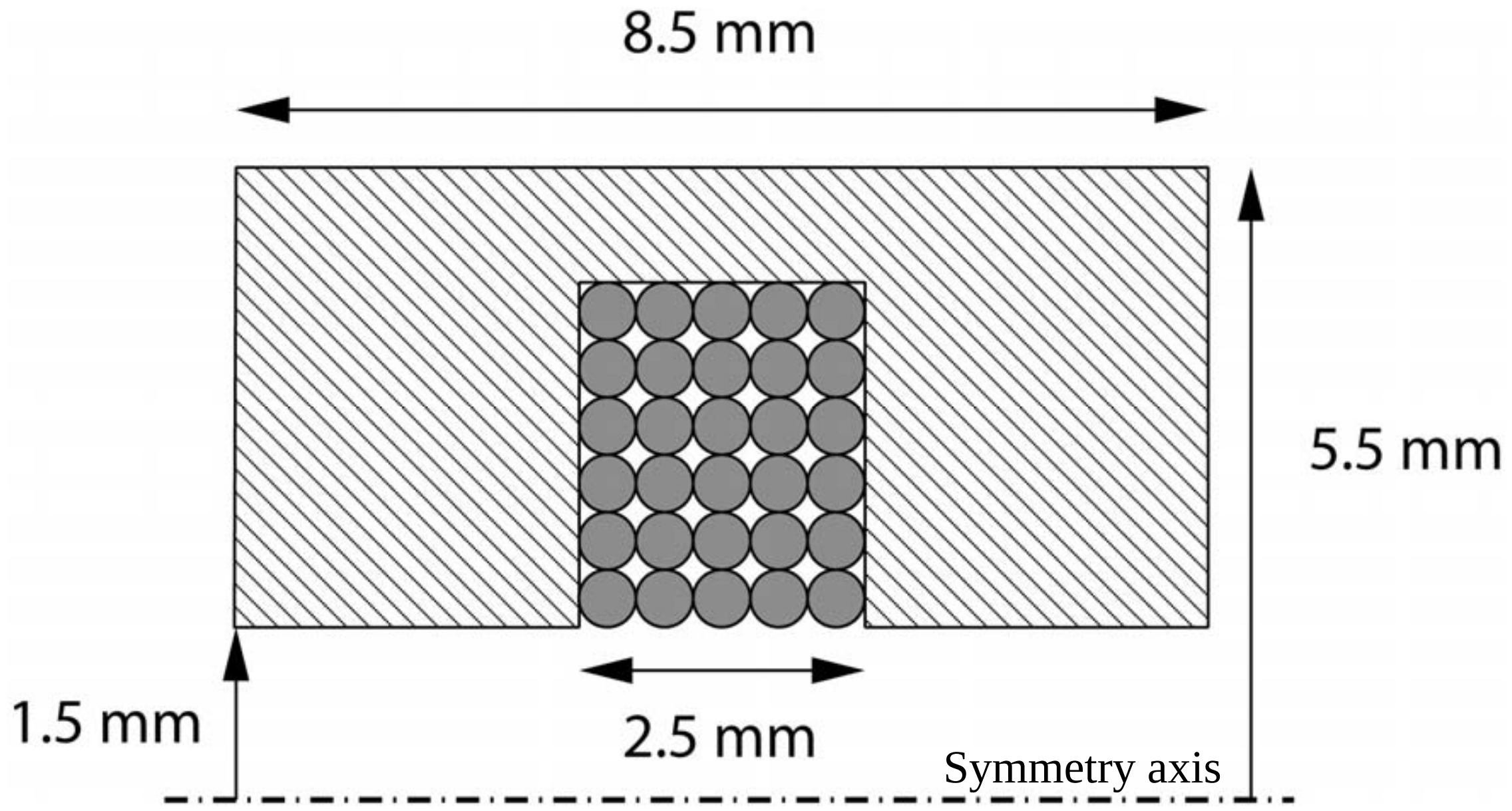

**Figure 1:** Cross-sectional view of our electromagnetic coil. The copper windings are enclosed in a Permendur shell.

## Electromagnetic coil design

The electromagnetic coil design is inspired by the solenoid valve used to pulse the supersonic nozzle. The cross section view of the coil is shown in Figure 1. It has 30 turns (5×6 layers) of 0.5 mm diameter copper wire and a 3 mm diameter bore. The copper windings are enclosed in a Permendur alloy shell. Permendur has a saturation magnetization of 2.3 Tesla[14] and allows us to achieve the highest ratio between magnetic flux and current density. We present the calculated axial and transverse magnetic fields of our coil in Figure 2. We use non-linear finite element analysis to take the magnetic saturation of Permendur into account.

According to our estimations a current density of $1\cdot10^9$ Amp/m$^2$, corresponding to 200 Amps, will produce a peak magnetic flux density of 2.1 Tesla.

The transverse magnetic potential is well approximated by a parabolic potential with a depth of 0.4 Tesla (Figure 2 b). The transverse field component focuses and guides the low field seekers during the slowing process by forming an effective magnetic guide.

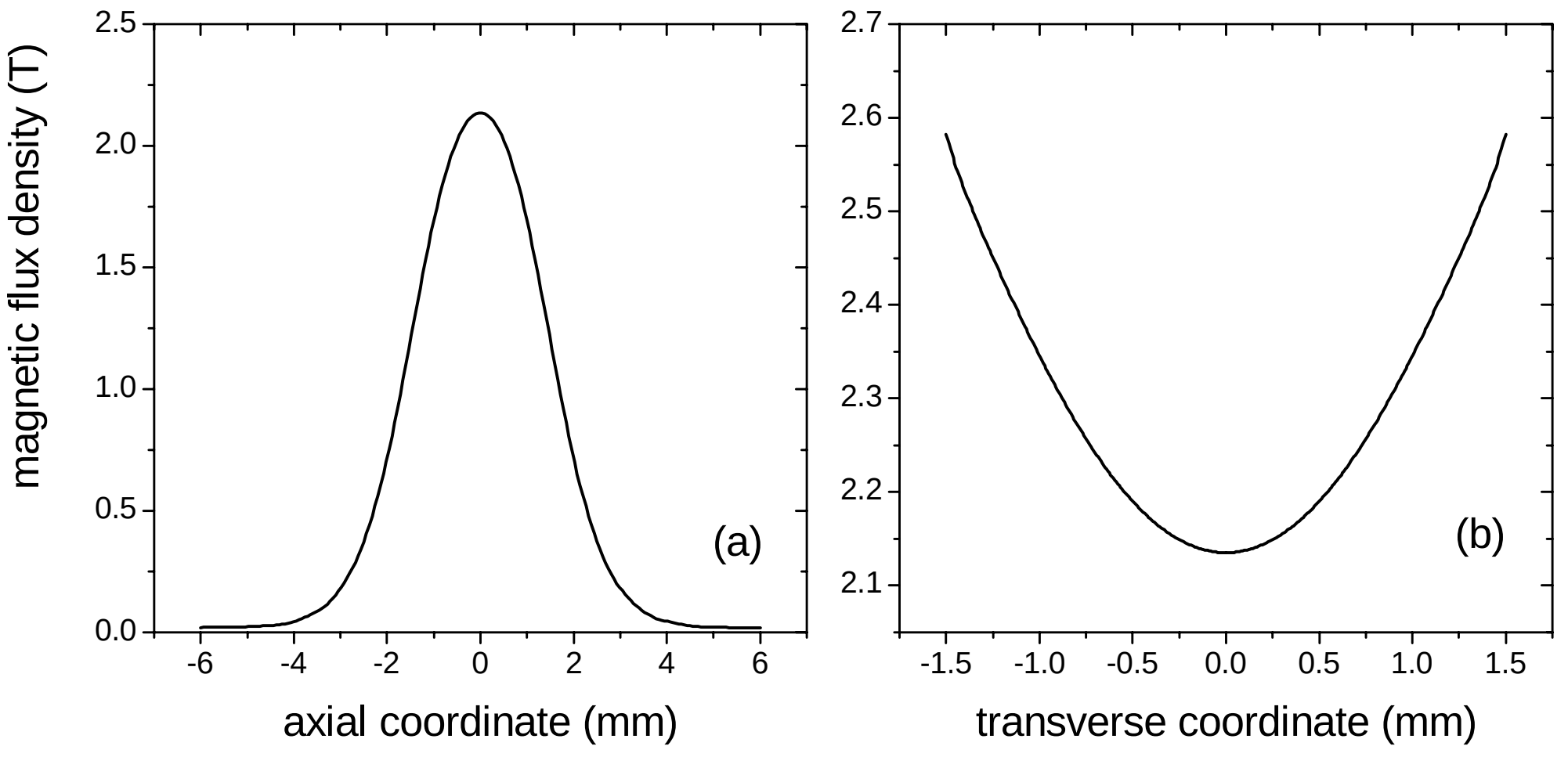

**Figure 2:** The axial (a) and transverse (b) components of the magnetic flux density at the center of the coil as calculated for the current density of $10^9$ A/m$^2$.

## Magnetic slowing simulation

Molecular oxygen is one of a few molecules that have a permanent magnetic moment in the ground state. It can be introduced into the supersonic beam by mixing it with a heavier carrier gas such as Xenon. The ground electronic state of molecular oxygen is a paramagnetic triplet (S=1) state, $^3\Sigma_g^+$. High intensity magnetic fields split the ground state oxygen into the three magnetic sub-level states with spin projections $M_s$ = 1, 0 and -1 (within Paschen-Back approximation). The low field seeker, $M_s$=1, atoms are slowed by the pulsed magnetic fields while Ms=0 atoms remain unaffected and Ms=-1 atoms are defocused.

To simulate the slowing of molecular oxygen, we numerically integrate the classical equations of motion using the magnetic fields obtained via finite element analysis. We model the electromagnetic coil switching as an exponential function having a time constant of 7 μsec. The simulated oxygen beam has an initial mean velocity 250 m/s with a standard deviation of 25 m/s. Our simulations only include those atoms within two standard deviations of the mean velocity. The standard deviation of the beam divergence angle is 0.1 rad and the standard deviation of the nozzle opening time is taken to be 15 μsec(corresponding to 35 μsec  FWHM pulse).

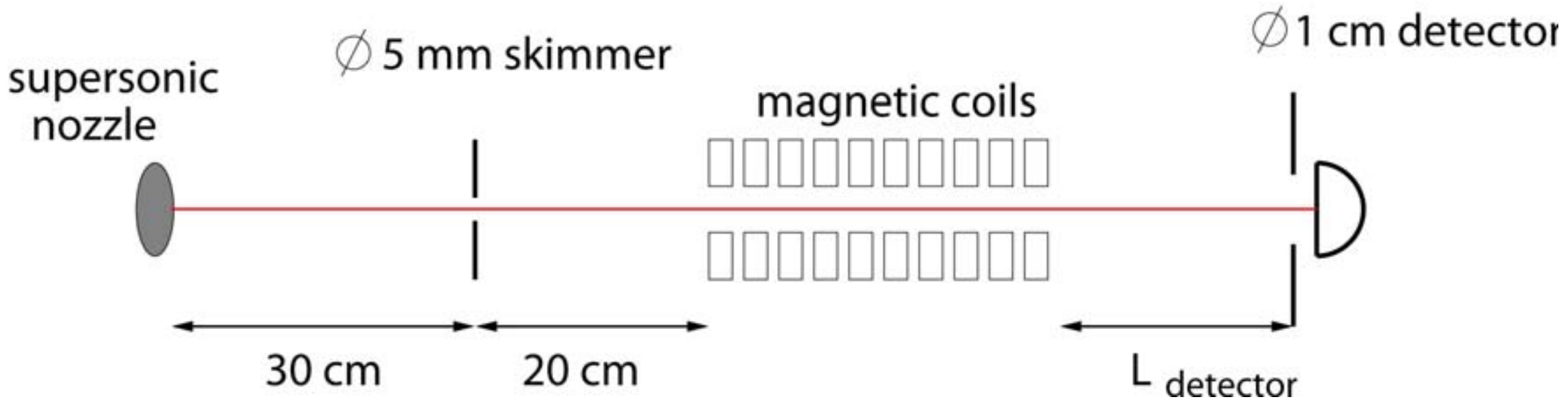

**Figure 3:** A schematic drawing of the magnetic slowing apparatus; the distance between the last coil and the detector, $L_{detector}$, is varied according to the number of the coils.

A schematic drawing of our simulated magnetic slower setup is presented in Figure 3. The distance between the supersonic nozzle and the first coil is 0.5 meters, the distance between coils (center-to-center) is 14.1 mm, and the distance from the last coil to the detector is 0.5 m, 0.25 m, and 0.05 m for 20, 48 and 54 coils respectively. We choose the timing of our coils such that the molecules with an initial velocity of 250 m/s lose the largest amount of kinetic energy per stage and are thus slowed the most effectively. We present the calculated arrival time distributions at the 1 cm diameter detector in Figure 4. The plots include arrival time calculations for all three possible $M_s$ states and show that the mean velocity of oxygen can be reduced to 50 m/s after 54 slowing stages.  The effective slowing is limited to the molecules that are within ±2 m/s of the mean velocity and only about 1 % of the detected molecules are slowed.

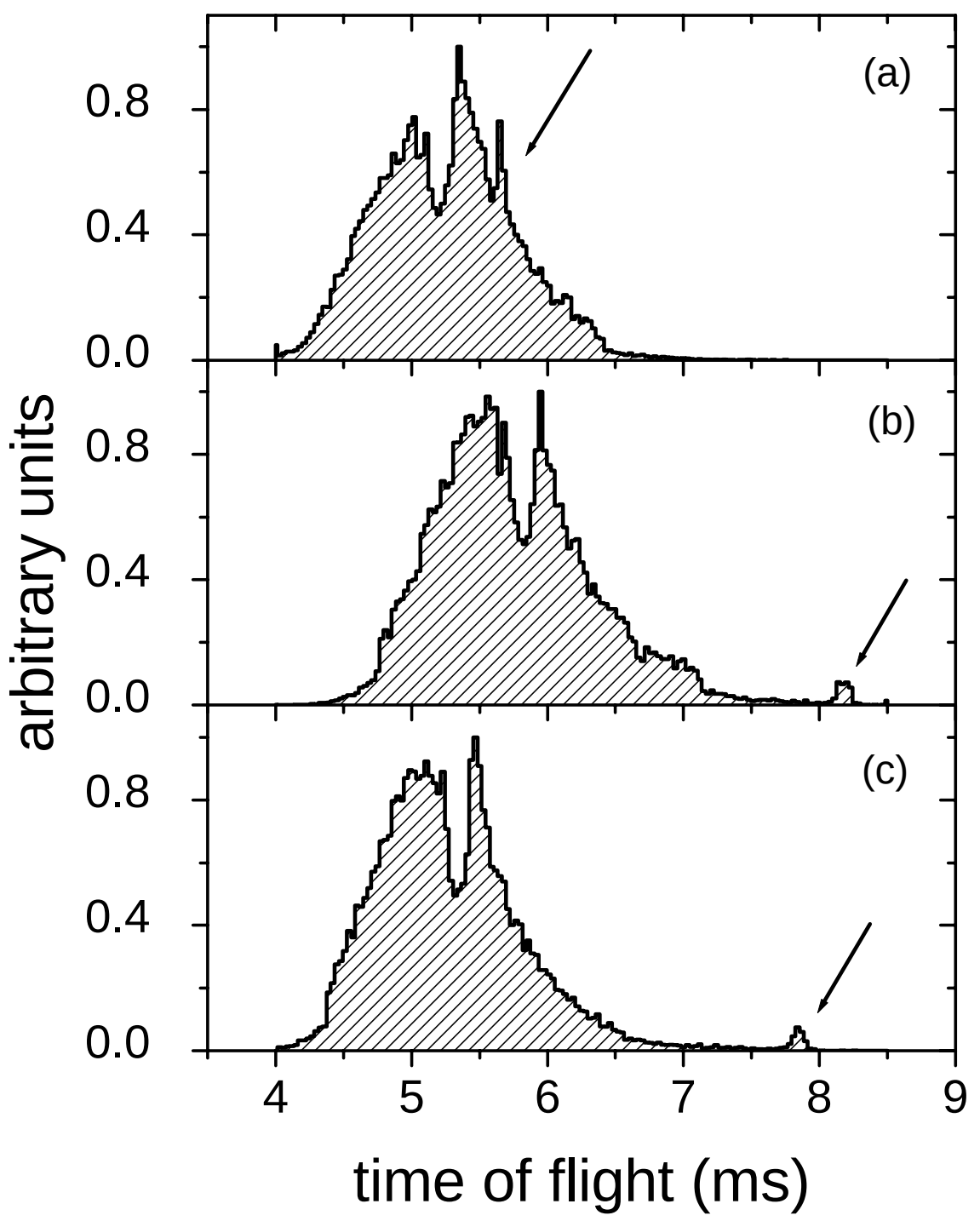

**Figure 4:** The results of time-of-flight simulations for 20 (a), 48 (b) and 54 (c) slowing coils.

## Moving magnetic trap

In principle, our proposed magnetic decelerator would be able to stop the atomic beam and enable atom transfer into a magnetic trap. A similar process has been demonstrated to produce electrostatically trapped polar molecules. This trapping technique is a two stage process that requires deceleration before trapping.

We propose to combine trapping *with* deceleration. Cold and fast atoms from the supersonic beam would be trapped in a co-moving, decelerating magnetic trap. The initial trap velocity is equal to the atomic beam velocity. Since the beam temperature in the moving frame is very low (under 100 mK), we can trap a large number of atoms. The number of atoms that "survive" the slowing process depends on the deceleration value.

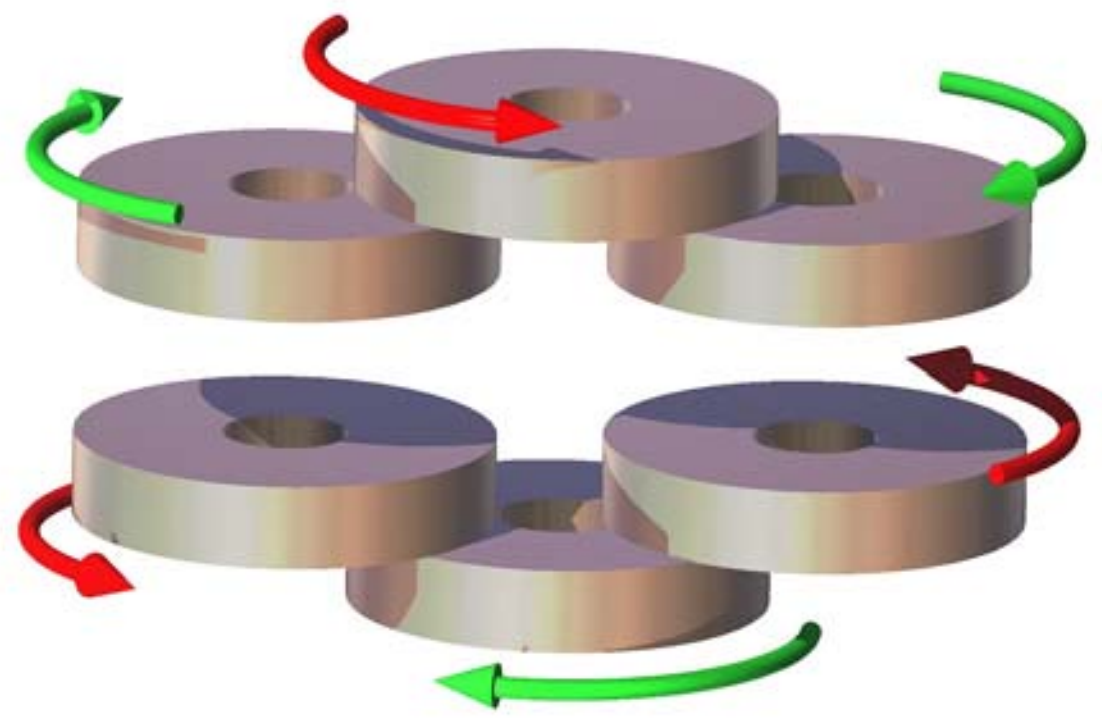

**Figure 5** Schematic drawing of three anti-Helmholz coil pairs that are sequentially activated to create a moving magnetic trap. The direction of the current in each coil is represented by an arrow.

Our proposed electromagnetic coil configuration for the adiabatic slowing process (see Figure 5) is similar to the setup used by Greiner et al.[15] to transfer laser cooled atoms over macroscopic distances. In our case the unit cell consists of three anti-Helmholz coil pairs that create a moving 1mm$^3$, 0.4 Tesla deep magnetic trap. The magnetic field minimum can be translated by changing the current ratio between two of the anti-Helmholz coil pairs.

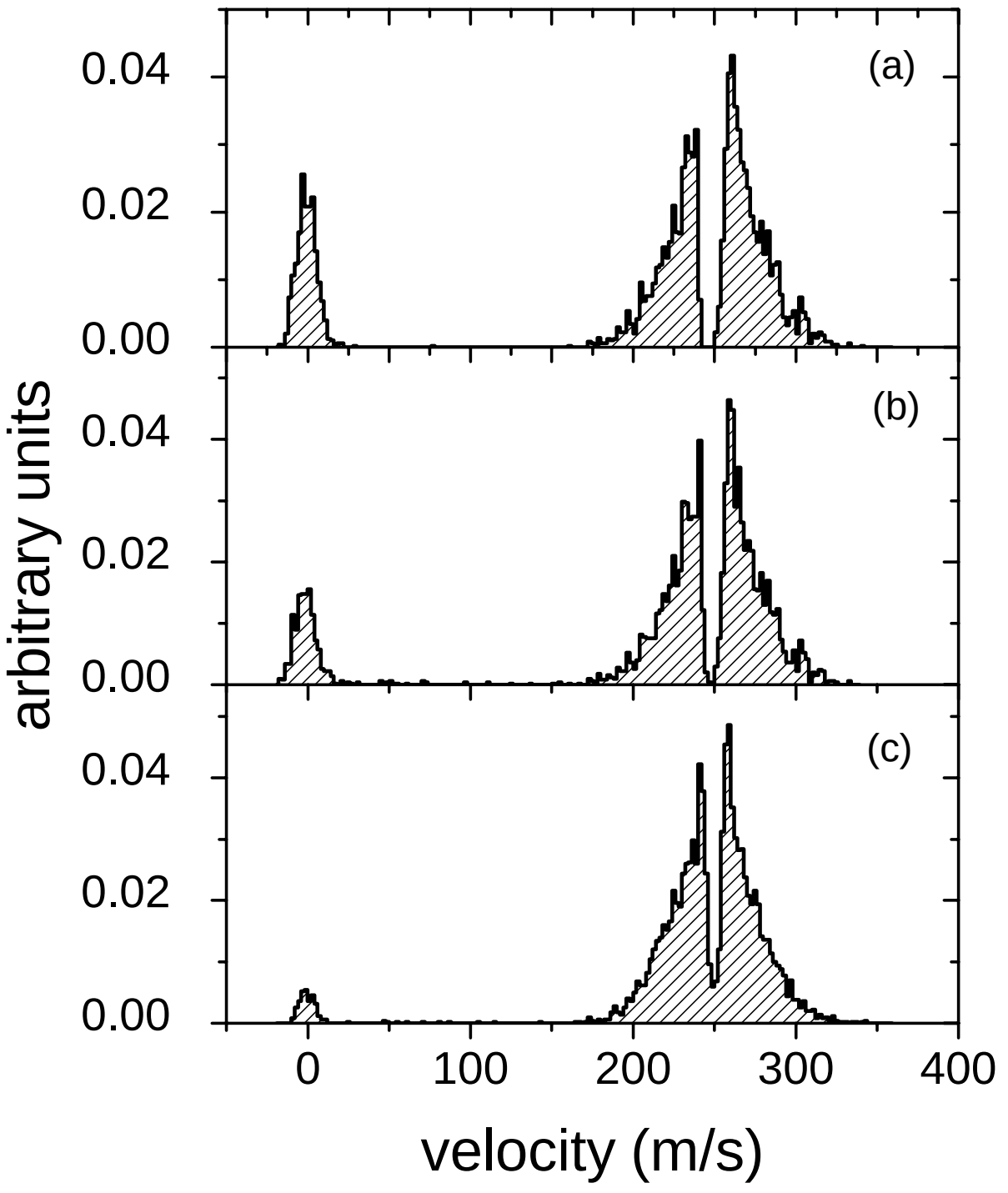

**Figure 6** Simulated velocity distributions of adiabatically stopped $O_2$ molecules in the $M_s$=1 state. The acceleration values are -4·10³ m/s² (a), -1·10⁴ m/s² (b), and -4·10⁴ m/s² (c).

We simulate the adiabatic slowing process for the case of $M_s$=+1 state of molecular oxygen. We use the same supersonic beam properties as in the previous section, however our calculations here are 1D. We switch the magnetic trap around the atoms instantaneously and slow it down with a constant deceleration. The resulting velocity distributions are shown in Figure 6. As expected, the number of slowed molecules depends on the deceleration. The trapping efficiency is about 5 % in the case of the largest acceleration of -40,000 m/sec$^2$. The total length of such an adiabatic decelerator device would be about 80 cm.

## Conclusions and Future Directions

In this paper we propose a novel deceleration method to slow paramagnetic atoms or molecules using pulsed magnetic fields. Our simulations predict that the mean velocity of molecular oxygen can be reduced from 250 m/sec to 50 m/sec in 54 stages, which would make a magnetic decelerator 76 cm in length. We also propose a method to adiabatically slow trapped particles to zero mean velocity by decelerating a moving magnetic trap formed by a series of anti-Helmholz coil pairs. More generally, the realization of a magnetic slower would provide a way to create cold atomic beams as well as trapped atoms for many species in the periodic table that until now could not be controlled.

We acknowledge discussions with Robert Hebner and the Center for Electromechanics at The University of Texas at Austin. We also thank Isaac Chavez for technical assistance. This work was supported by the Army Research Office, the R. A. Welch Foundation, and the Sid W. Richardson Foundation.

[12] W. Gerlach and O.Stern, Zeit. Phys. **9**, 349 (1922)

[13] L. D. Landau and E. M. Lifshitz, *Quantum Mechanics* (Pergamon, Oxford, 1965).

[14] R. A. Early, J. K. Cobb, J. E. Oijala, Particle Accelerator Conference, 1989. 'Accelerator Science and Technology'., Proceedings of the 1989 IEEE, 351 vol. 1 (1989)

[15] M. Greiner, I. Bloch, T. W. Hänsch, and T. Esslinger, Phys. Phys. Rev. A **63**, 031401 (2001)